\documentclass[aps,prd,superscriptaddress,preprintnumbers]{revtex4}

\preprint{To appear in the proceedings of DSPIN-19}

\usepackage{pdfsync}
\usepackage{mathptmx}
\usepackage[utf8]{inputenc}
\usepackage[T1]{fontenc}
\usepackage{slashed}
\usepackage{times}
\usepackage{amssymb,amsfonts,amsmath,amsthm}
\usepackage{dsfont,bbm}
\usepackage{dcolumn}
\usepackage{epsf}
\usepackage{graphicx}
\usepackage[caption=false]{subfig}
\usepackage{dsfont}
\usepackage{slashed}
\usepackage[active]{srcltx}
\usepackage[usenames]{color}

\begin{document}

\title{Light-Cone Sum Rules for Gravitational Form Factors}

\author{I.~V.~Anikin}

\affiliation{Bogoliubov Laboratory of Theoretical Physics, JINR,
             141980 Dubna, Russia}

\email{anikin@theor.jinr.ru}

\begin{abstract}
We outline the developed approach within the light-cone sum rules at the leading order for calculation of
the gravitational form factors related to the nucleon valence quark combinations.
The predictions for the gravitational form factor $\mathds{D}(t)$ ($D$-term contributions) have been presented.
\end{abstract}
\maketitle

\section{Introduction}
\label{Intro}

The hadron matrix element of energy-momentum tensor (EMT) can provide information
on fundamental characteristics of particles such as mass and spin \cite{Polyakov:2002yz, Polyakov:2018zvc, Polyakov:2018exb, Belitsky:2005qn, Diehl:2003ny, Ji:1996ek, Balitsky:1997rs}.
The special attantions have been paied in order to study the different relations between the gravitational form factors $\mathds{A}(t)$ and $\mathds{B}(t)$, parameterizing
the hadron matrix element of EMT, and the Mellin moment of generalized parton distributions
$H(x, \xi, t)$ and $E(x, \xi, t)$ as known as Ji's sum rules
(see, for example, \cite{Belitsky:2005qn, Diehl:2003ny}).
Moreover,
one of the EMT form factors is associated with the $D$-term \cite{Kivel:2000fg} which is probably
the last unknown fundamental hadron characteristic determining the spatial deformations and
defining the mechanical properties of hadrons
\cite{Polyakov:2002yz, Polyakov:2018zvc, Polyakov:2018exb, Teryaev:2013qba, Pasquini:2014vua, Lorce:2018egm, Wakamatsu:2007uc}.

As snown in \cite{Anikin:2019kwi, Anikin:2019ufr},
the light-cone sum rules approach (LCSRs) \cite{Anikin:2013aka, Anikin:2015ita, Anikin:2016teg} which has been developed to compute the different nucleon electromagnetic form factors can be adopted for the study of the different gravitational form factors.

In the present proccedings, we are outlining the main features of our LCSR approach to calcultions and/or estimations
of  the nucleon gravitational form factors.
Our final predictions demonstrate reasonably good agreements with the first experimental data \cite{Hagler:2007xi, Burkert:2018bqq},
the chiral quark-soliton and Skyrme model results for $D$-term contributions \cite{Polyakov:2018zvc}.

\section{Energy-momentum tensor}
\label{Subsec:emt-tw}

We begin with the quark contribution to the Belinfant\'{e} improved energy-momentum tensor
which is identical to the local geometrical twist-$2$ operator. We have
$
2\,\Theta^{\mu\nu}_q(0)= i\,{\cal R}^{\mu\nu}_{\tau=2}
$
and it can be expressed through the non-local operator as (cf. \cite{Balitsky:1987bk})
\begin{eqnarray}
\label{emt-2}
-2i\,\Theta^{\mu\nu}_q(0)&=&\lim_{y\to 0}\frac{\partial}{\partial y_\nu}\, \int\limits_{0}^{1}du\,\frac{\partial}{\partial y_\mu}
\big[ \bar\psi(0)\, \hat y \,[0\,;\,uy]_A\, \psi(uy)  \big]
-\text{(trace)}.
\end{eqnarray}

Let us also introduce the auxiliary geometrical twist-2 operator defined as
\begin{eqnarray}
\label{tw-2-2}
&&\widetilde{\cal R}^{\mu\nu}_{\tau=2}=\lim_{y\to 0}\frac{\partial}{\partial y_\nu}\, \int\limits_{0}^{1}du\,\frac{\partial}{\partial y_\mu}
\big[ \bar\psi(0)\, y_\alpha\gamma^{\alpha,+} \,[0\,;\,uy]_A\, \psi(uy)  \big]
\end{eqnarray}
which actually differs from ${\cal R}^{\mu\nu}_{\tau=2}$ but
$\widetilde{\cal R}^{+ +}_{\tau=2} = {\cal R}^{+ +}_{\tau=2}$.
The reason for the introduction of $\widetilde{\cal R}^{\mu\nu}_{\tau=2}$ is the following:
it is well-known \cite{Braun:2006hz} that for the electromagnetic form factors the LCSRs can be established self-consistently only
for the plus light-cone projection of electromagnetic current, $J_{em}^+(0)$, which corresponds to the twist-$2$ operator combination of
the current.
Indeed, in the case of $J_{em}^-(0)$ the leading pole term, which corresponds to the {\it l.h.s.} of the sum rules, takes the form
\begin{eqnarray}
\label{minus-J}
\frac{\lambda_1}{m^2_N-P^{\prime\, 2}}
\Big(
m_N \big[ c_1 \, F_1(Q^2) + c_2\, F_2(Q^2)\big] +
\hat q_\perp \big[ c_3 \, F_1(Q^2) + c_4\, F_2(Q^2) \big]
\Big)N^+(P),
\end{eqnarray}
where $c_i$ are the known functions of the invariants.
In order to derive the sum rules the combination at $m_N \hat 1$ should relate to the ${\cal A}$-type of the amplitude, {\it i.e.}
${\cal A}^{-}_{em}(x_i; \Delta^2, P^{\prime\, 2})$, while
the combination at $\hat q_\perp$ -- to the ${\cal B}$-type of the amplitude, {\it i.e.}
${\cal B}^{-}_{em}(x_i; \Delta^2, P^{\prime\, 2})$ \cite{Anikin:2013aka}.
From (\ref{minus-J}), we can see that,
in contrast to the plus light-cone projection of electromagnetic current ,
each of the independent tensor structures $m_N \hat 1$ and $\hat q_\perp$ includes both $F_1(Q^2)$ and $F_2(Q^2)$.
 If we now consider the totally collinear case for the quark operators which form the corresponding correlator,
the ${\cal B}$-type of the leading order amplitude for the minus light-cone projection of electromagnetic current disappears.
Hence, in this approximation the form factors $F_1(Q^2)$ and $F_2(Q^2)$
are not independent ones contradicting the natural assumptions.

Having kept only the plus light-cone projection of the non-local operator,
we are able to develop LCSRs for the gravitational form factors by means of
appropriate adoption of our preceding calculations implemented for
the electromagnetic form factors \cite{Anikin:2013aka} (see eqns.~(\ref{emt-2}) and (\ref{tw-2-2})).
As a result, we are limited by the bilinear quark combinations with the spin projection $s_a=+1$,
{\it  i.e.} we deal with the collinear twist-$2$ quark combination $[\bar\psi_+\, \psi_+]$ of ${\cal R}^{\mu\nu}_{\tau=2}$.
For example, excluding the trivial case of the plus-plus projection,
we consider ${\cal R}^{+-}_{\tau=2}$ which is
\begin{eqnarray}
\label{pmR}
{\cal R}^{+-}_{\tau=2}=
\frac{1}{2} \Big( \bar\psi(0)\, \gamma^{+}\vec{\cal D}^{-}\,\psi(0) +  \bar\psi(0)\, \gamma^{-}\vec{\cal D}^{+}\,\psi(0) \Big)
=\widetilde{\cal R}^{+-}_{\tau=2}  + \frac{1}{2} \bar\psi(0)\, \gamma^{-}\vec{\cal D}^{+}\,\psi(0).
\end{eqnarray}
Here, the first term with $[\bar\psi_+\, \psi_+]$-combination is traded for $\widetilde{\cal R}^{+-}_{\tau=2}$ while
the second term with $[\bar\psi_-\, \psi_-]-$combination is kept intact and it is beyond the direct computations within our approach.
Despite, the $[\bar\psi_-\, \psi_-]-$ and $[\bar\psi_+\, \psi_-]-$contributions to $\Theta^{+-}$ and $\Theta^{+\perp}$
remain unavailable for the explicit LCSRs calculations,
because they are given by ``bad'' projections $J_{em}^-(0)$ and $J_{em}^\perp(0)$.
However, we can still make several estimations for this kind of contributions.

\section{Gravitational Form Factors within the LO LCSR}
\label{Sec:GFF-LO}

For the quark contribution, the gravitational form factors parametrize
the hadron matrix element of the Belinfant\'{e} improved energy-momentum tensor operator as \cite{Polyakov:2018exb}
\begin{eqnarray}
\label{EMT-me-1}
\langle P^\prime | \Theta^{(q)}_{\mu\nu}(0) | P \rangle =
\bar N(P^\prime) \Big[ \mathds{A}(t)\frac{\overline{P}_\mu\overline{P}_\nu}{m_N} +
i\mathds{J}(t)\frac{\overline{P}_{\left\{\mu\right.}\sigma_{\left.\nu\right\}\Delta}}{m_N} +
\mathds{D}(t)\frac{\Delta_\mu\Delta_\nu -g_{\mu\nu} \Delta^2}{4m_N} + g_{\mu\nu} m_N \overline{\mathds{C}}(t)
\Big] N(P),
\end{eqnarray}
where $\mathds{J}(t)=( \mathds{A}(t) + \mathds{B}(t))/2$,
and
$\overline{P}=( P^\prime + P)/2$ together with $\Delta=P^\prime - P$, $\Delta^2=-t$.
The preponderance of the Belinfant\'{e}-improved EMT is that it includes the contribution from the spin momentum tensor.

As demonstrated in \cite{Anikin:2019kwi}, the projection ${\cal R}^{++}_{\tau=2}$ is fully enough to compute
only the form factors $\mathds{A}(t)$ and $\mathds{B}(t)$.
With respect to the form factor $\overline{\mathds{C}}(t)$, with the help of the QCD equations of motion
the corresponding operator can be re-expressed through the matrix elements of the quark-gluon operator,
focusing on for the quark contribution, or the gluon-gluon operator, working with the gluon contribution \cite{Tanaka:2018wea}.
Therefore, the calculation of $\overline{\mathds{C}}(t)$ is beyond the leading order.
For calculation of the form factor $\mathds{D}(t)$, a projection ${\cal R}^{+-}_{\tau=2}$ is necessary.
At the same time, our exact computations are restricted by the ``good'' $[\bar\psi_+ \, \psi_+]-$combinations.
Hence, instead of the projections of ${\cal R}^{\mu\nu}_{\tau=2}$
we work with the projections of $\widetilde{\cal R}^{\mu\nu}_{\tau=2}$.
Also, we propose a reliable recipe for estimations of the ``bad'' $[\bar\psi_- \, \psi_-]-$combinations.

In analogy with the nucleon electromagnetic form factors \cite{Anikin:2013aka, Anikin:2015ita, Anikin:2016teg},
we define the amplitude which corresponds to the hadron matrix element of the energy-momentum tensor as
\begin{eqnarray}
\label{gravi-amp-1}
&&\hspace{-0.4cm}T^{\mu\nu}_{[\bar\psi_+ \psi_+]}(P,\Delta)\equiv\langle P^\prime| \widetilde{\cal R}^{\mu\nu}_{\tau=2} | P\rangle
=\lim_{y\to 0}\frac{\partial}{\partial y_\nu}\, \int\limits_{0}^{1}du\,\frac{\partial}{\partial w_\mu}
\int(d^4z) e^{-i\Delta\cdot z} \langle 0| T \eta(0)
\nonumber\\
&&\hspace{-0.4cm}
\times\big[ \bar\psi(w+z)\, w^\alpha\gamma^{\alpha,+}\,
[w+z\,;\,-w+z]_A\, \psi(-w+z) \big]| P\rangle
\end{eqnarray}
where $w=uy$ and $
\eta(0)=\varepsilon^{ijk}\big[ u^i(0) C \gamma_\alpha u^j(0)\big] \gamma_5 \gamma_\alpha d^k(0)$.

By the straightforward calculation of (\ref{gravi-amp-1}) with the DA parametrizations
(see for example \cite{Anikin:2013aka}), one can see
that in order to calculate the corresponding gravitational form factors
we merely have to weight the electromagnetic form factors with the certain tensor structure.
Indeed, the exact amplitude for $d$-quark contribution reads
\begin{eqnarray}
\label{Adopt-2}
T^{\mu\nu\,(d)}_{[\bar\psi_+ \psi_+]}(P,\Delta) =
\frac{1}{8} \int {\cal D}x_i
\Big( 2p_{\mu} \big[ 2x_1 P + \Delta \big]_{\nu} + (\mu\leftrightarrow\nu) \Big)
T^{(d)\,+}_{em}(x_i; P,\Delta),
\end{eqnarray}
where
\begin{eqnarray}
\label{Adopt-2-2}
T^{(q)\,+}_{em}(x_i; P,\Delta) =
 \Big\{
m_N\, {\cal A}^{+\,(q)}_{em}(x_i; \Delta^2, P^{\prime\, 2}) +
\hat\Delta_\perp \,{\cal B}^{+\,(q)}_{em}(x_i; \Delta^2, P^{\prime\, 2})\Big\} N^+(P).
\end{eqnarray}
and ${\cal A}^{+\,(q)}_{em}$, ${\cal B}^{+\,(q)}_{em}$ have been taken from \cite{Anikin:2013aka}.
Therefore, we obtain that the tensor structure of the amplitude (\ref{gravi-amp-1})
can entirely be defined by the tensor
$ 4 p_{\left\{ \mu\right.} \big[ 2x_i P + \Delta \big]_{\left.\nu\right\}}$.

\subsection{Gravitational form factors $ \mathds{A}(t)$ and $ \mathds{B}(t)$}
\label{Subsec:plus-plus-proj}

First, we dwell on the $(+,\,+)$ light-cone projection of the amplitude given by
\begin{eqnarray}
n_\mu\,n_\nu \, T^{\mu\nu}(P,\Delta)=T^{++}(P,\Delta)=
\Big[ m_N\, \mathcal{A}^{++}(P,\Delta) + \hat\Delta_\perp \mathcal{B}^{++}(P,\Delta)\Big] N^+(P).
\end{eqnarray}
Making use of the Borel transforms one obtains the sum rules
\begin{eqnarray}
\label{A-B-ff}
\mathds{A}(t) = \frac{1}{2} \int \hat d\mu(s)\Big\{\mathcal{A}^{++}_{{\rm QCD}}(\Delta^2,s) \Big\}, \quad
\mathds{B}(t) = -\int \hat d\mu(s)\Big\{\mathcal{B}^{++}_{{\rm QCD}}(\Delta^2,s) \Big\}.
\end{eqnarray}
where
\begin{eqnarray}
\label{DR-short}
\int \hat d\mu(s)\big\{\mathcal{F} \big\}
=\frac{1}{\lambda_1\pi}\int_{0}^{s_0} ds\, e^{(m_N^2-s)/M^2}\, \text{Im} \big\{\mathcal{F}(s,t) \big\}.
\end{eqnarray}

Following our strategy \cite{Anikin:2019kwi}, since the gravitational form factors $\mathds{A}(t)$ and $\mathds{B}(t)$
are not reachable for a small region of $t$, we implement the approximate fits for these form factors in the region of large $t$ and
make an analytical continuation of the obtained fitting functions to the region of small $t$ afterwards.
The exact results for $\mathds{A}(t)$ and $\mathds{B}(t)$ can be found in \cite{Anikin:2019kwi}.
Here, we are limited by the presentation of the gravitational form factors $\mathds{A}(t)$ and $\mathds{B}(t)$
in the form of the multipole functions as (both for ABO2-parametrization of \cite{Anikin:2013aka})
\begin{eqnarray}
\label{A-fit}
\mathds{A}^{\text{fit}}(t) = \frac{\kappa}{(1+a\, t)^b},\quad
\kappa=1.01,\quad a=0.7\,\text{GeV}^{-2},\quad b= 2.95,
\mathds{A}^{\text{fit}\,\prime}(0)=-2.1,
\end{eqnarray}
and
\begin{eqnarray}
\label{B-fit}
\mathds{B}^{\text{fit}}(t) =
\frac{k\, t}{(1+ c\, t)^d}, \quad
k=0.073, \quad c=0.45\,\text{GeV}^{-2}, \quad d=4.1\,.
\end{eqnarray}

\subsection{Gravitational form factor $ \mathds{D}(t)$}
\label{Subsec:plus-minus-proj}

As demonstrated in detail, see \cite{Anikin:2019kwi}, the gravitational form factor $\mathds{D}(t)$ can reasonably
be estimated as
\begin{eqnarray}
\label{Dfull-Est}
\mathds{D}(t) \approx
 - \int \hat d\mu(s) \Big\{
\frac{1}{2}\mathcal{B}^{++} +
\frac{m^2_N}{t} \mathcal{A}^{++} \Big\}.
\end{eqnarray}
This means that the form factor $\mathds{D}(t)$, as a independent function of $t$,
behaves similary to a combination of the form factors $\mathds{A}(t)$ and  $\mathds{B}(t)$.

\begin{figure}[h]
\includegraphics[width=16pc]{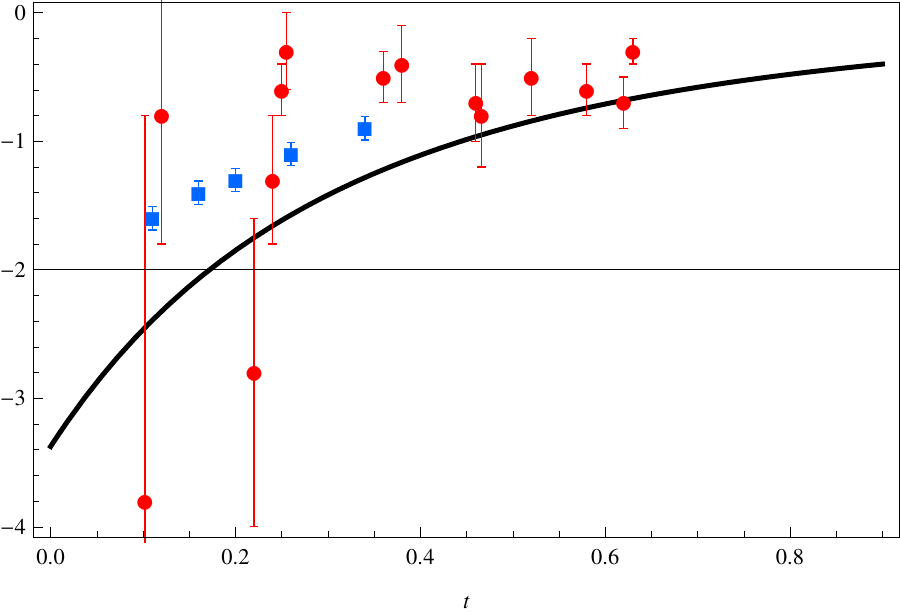}\hspace{2pc}%
\begin{minipage}[b]{14pc}\caption{The gravitational form factor $\frac{5}{4}\mathds{D}^{\text{fit}}(t)$
which is analytically continued to the small region of $t$.
The experimental data on the nucleon $D$-term form factor:
the blue squares from the Jefferson Lab \cite{Burkert:2018bqq}, the red bullets from
the Lattice QCD \cite{Hagler:2007xi}.}
\label{Fig-D-fit}
\end{minipage}
\end{figure}

As above, the estimated $\mathds{D}(t)$ can be approximated by (cf. \cite{Tanaka:2018wea})
\begin{eqnarray}
\label{Dtot-fit}
\mathds{D}^{\text{fit}}(t) = \frac{D}{(1+e\, t)^n}, \quad
D=-2.5, \quad e=0.93\,\text{GeV}^{-2}, \quad n= 3.4
\end{eqnarray}
In the region where $t$ is small, for the full set of LCSR-parametrizations \cite{Anikin:2013aka},
this approximating function matches, see Fig.~\ref{Fig-D-fit}, with both the first experimental data
\cite{Burkert:2018bqq,Hagler:2007xi}
and the results of the chiral quark-soliton and Skyrme models presented in \cite{Polyakov:2018zvc}.
Our function $\mathds{D}^{\text{fit}}(t)$ lies below than the quark-soliton prediction and above than the result of Skyrme
model.

\subsection{$(+, \perp)$ light-cone projections of the amplitude}
\label{Subsec:plus-perp-proj}

As the last step, for the sake of completeness, we present the $(+, \perp)$ light-cone projection of the amplitude,
$T^{+\perp}(P,\Delta)$. We have
\begin{eqnarray}
\label{Pol-sr-3}
\frac{\lambda_1 m_N \Delta^2_\perp}{m^2_N - P^{\prime\,2}}
 \mathds{A}(t) N^+(P) =
m_N\Big(
\bar T^{+\perp}_{1\,[\bar\psi_+ \psi_+]} + \bar T^{+\perp}_{1\,[\bar\psi_+ \psi_-]}
\Big) N^+(P)
\nonumber\\
\end{eqnarray}
where
\begin{eqnarray}
\label{Pol-sr-3-2}
m_N \Big(
\bar T^{+\perp}_{1\,[\bar\psi_+ \psi_+]} + \bar T^{+\perp}_{1\,[\bar\psi_- \psi_-]}
\Big)=m_N \bar T^{+\perp}_{1\,[\bar\psi_- \psi_-]} +
\frac{m_N }{\pi}\int_{0}^{s_0} \frac{ds}{s-P^{\prime\,2}} \text{Im}\Big\{
\frac{\Delta^2_\perp}{4}\mathcal{A}^+_{em}\Big\},
\end{eqnarray}
and
\begin{eqnarray}
\label{Pol-sr-4}
\frac{\lambda_1}{m^2_N - P^{\prime\,2}}
\Big( P\cdot\Delta \mathds{J}(t) +
\frac{\Delta^2_\perp}{4} \big[\mathds{A}(t)-\mathds{B}(t)\big]
\Big) \hat\Delta_\perp N^+(P)
=\hat\Delta_\perp\Big(
\bar T^{+\perp}_{4\,[\bar\psi_+ \psi_+]} + \bar T^{+\perp}_{4\,[\bar\psi_+ \psi_-]}
\Big) N^+(P)
\end{eqnarray}
where
\begin{eqnarray}
\label{Pol-sr-4-2}
\hat\Delta_\perp \Big(
\bar T^{+\perp}_{4\,[\bar\psi_+ \psi_+]} + \bar T^{+\perp}_{4\,[\bar\psi_- \psi_-]}
\Big)=\hat\Delta_\perp \bar T^{+\perp}_{4\,[\bar\psi_- \psi_-]} +
\frac{\hat\Delta_\perp }{\pi}\int_{0}^{s_0} \frac{ds}{s-P^{\prime\,2}} \text{Im}\Big\{
\frac{\Delta^2_\perp}{4}\mathcal{B}^+_{em}\Big\}.
\end{eqnarray}
Using the Borel transforms and eqn.~(\ref{A-B-ff}), we derive the following representations
\begin{eqnarray}
\label{F-1}
&&
\int \hat d\mu(s) \Big\{ \frac{1}{2}\mathcal{A}^{++} \Big\}
= \int \hat d\mu(s) \Big\{ \frac{1}{4}\mathcal{A}^{+}_{em} \Big\}+ \tilde T^{+\perp}_{1\,[\bar\psi_+ \psi_-]}, \,
\\
&&
\int \hat d\mu(s) \Big\{ \frac{1}{2}\mathcal{B}^{++} \Big\} =
\int \hat d\mu(s) \Big\{ \frac{1}{4}\mathcal{B}^{+}_{em} \Big\} + \tilde T^{+\perp}_{4\,[\bar\psi_+ \psi_-]}.
\end{eqnarray}
These sum rules allow us to calculate the contributions of $\tilde T^{+\perp}_{1, 4\, [\bar\psi_+ \psi_-]}$ explicitly.

\section{Conclusions and discussions}
\label{Sec:Conclusions}

As well-known (see, for example, \cite{Balitsky:1986st, Balitsky:1989ry, Braun:2006hz, Braun:2001tj}),
the approaches based on the light-cone sum rules are very attractive because the soft contributions are calculated in terms of
the distribution amplitudes.

In this proceedings, we have outlined the calculations and/or estimations for
the gravitational form factors within the frame of the LCSRs techniques at the leading order.

We have demonstrated that the essential contributions to the gravitational form factors can be calculated thanks for
the suitable modification of the LCSRs approach developed for the electromagnetic form factors.
For the valence quark content of nucleon,
the gravitational form factors $\mathds{A}(t)$  and $\mathds{B}(t)$ are reachable for direct calculations
in the region of sufficiently large $t\gtrsim 1\, \text{GeV}^2$ where one can rely on the LCSRs.
For this region, the obtained form factors can be approximated by the corresponding multipole functions
$\mathds{A}^{\text{fit}}(t)$  and $\mathds{B}^{\text{fit}}(t)$ and, then, can analytically be continued to
the region of small $t$.

Also, we have presented the estimation for the valence quark contributions to
the gravitational form factor $\mathds{D}(t)$ for both the regions of large and small $t$.
Regarding the region of small $t$, where the first experimental data are available \cite{Hagler:2007xi, Burkert:2018bqq},
the estimated $\mathds{D}(t)$ has been approximated by the fitting multipole function
$\mathds{D}^{\text{fit}}(t)$ which, together with the corresponding gluon contributions, will allows us to calculate
a few quantities characterizing the ``mechanical'' properties of the nucleon (this work is now in progress).
Our results agree rather well with the experimental data and with the results
of the chiral quark-soliton and Skyrme models \cite{Polyakov:2018zvc}.


\end{document}